\documentclass[%
 aip,
 amsmath,amssymb,
 reprint,%
]{revtex4-1}

\usepackage{graphicx}
\usepackage{dcolumn}
\usepackage{bm}

\usepackage{siunitx}

\usepackage[utf8]{inputenc}
\usepackage[T1]{fontenc}
\usepackage{mathptmx}
\usepackage{etoolbox}

\usepackage{CJKutf8}
\makeatletter
\def\@email#1#2{%
 \endgroup
 \patchcmd{\titleblock@produce}
  {\frontmatter@RRAPformat}
  {\frontmatter@RRAPformat{\produce@RRAP{*#1\href{mailto:#2}{#2}}}\frontmatter@RRAPformat}
  {}{}
}%
\makeatother
\begin{document}
\begin{CJK*}{UTF8}{gbsn}

\preprint{AIP/123-QED}

\title{Simulation of Flagellated Bacteria Near a Solid Surface: Effects of Flagellar Morphology and Ionic Strength}
\author{Baopi Liu}
\email{bpliu@mail.bnu.edu.cn}
\affiliation{School of Physics, Ningxia University, Yinchuan, Ningxia 750021, China}
\affiliation{School of Advanced Interdisciplinary Studies, Ningxia University, Zhongwei, Ningxia 755000, China}
\affiliation{Complex Systems Division, Beijing Computational Science Research Center, Beijing 100193, China}

\author{Bowen Jin}
\affiliation{Huanjiang Laboratory, Zhuji, Zhejiang 311899, China}
\affiliation{State Key Laboratory of Fluid Power and Mechatronic Systems, Department of Mechanics, Zhejiang University, Hangzhou, Zhejinag 310027, China}

\author{Ning An}
\affiliation{School of Physics and Astronomy, Beijing Normal University, Beijing 100875, China}

\date{\today}
\begin{abstract}
This study systematically investigates the three-stage process of bacterial surface entrapment, characterized by its height and inclination angle. Initially, bacteria swim towards the surface at an approach velocity proportional to motor rotation frequency. Subsequently, the cotangent of the inclination angle decays exponentially with the product of the motor rotation frequency and time during reorientation. Finally, under the combined action of near-field hydrodynamic interactions and DLVO forces, bacteria reach a stable fixed point near the surface. Bacteria with left-handed chiral flagella exhibit a clockwise circular motion on the surface. The stable heights, inclination angles, and radii of curvature of these circular trajectories are collectively determined by the flagellar morphology and ionic strength of the electrolyte solution. Specifically, increasing the contour length of the flagellum reduces the stable inclination angle and increases the radius of curvature. In contrast, decreasing the ionic strength increases the stable height and radius of curvature, while also decreasing the stable inclination angle. Typically, the stable inclination angle falls within $(\pi/2,\pi)$, the stable height ranges from several nanometers to over one hundred nanometers, and the radius of curvature spans several to tens of micrometers. Our work explains the observed dispersion of the stable heights.
\end{abstract}
\maketitle
\end{CJK*}
\section{Introduction}
Many Bacteria swim in a fluid medium by generating propulsion through the rotation of flagella~\cite{Diluzio2005,Lauga2016}. Near surfaces, the interactions between the bacteria and the solid-liquid interface are significantly influenced by boundary conditions, which specify the slip velocity at the interface. Boundary conditions can be classified into three categories: no-slip, partial-slip, and perfect-slip~\cite{Lauga2007}. They play an important role in surface interactions and bacterial motility, thereby affecting processes such as bacterial adhesion and biofilm formation. Bacterial biofilms are complex communities of bacteria embedded in an extracellular polymeric substance matrix~\cite{Krsmanovic2021}. Biofilm formation is closely related to numerous infectious diseases and contamination of medical devices~\cite{Krsmanovic2021,Costerton1999,Kargar2014,Van2018,Carniello2018}. Therefore, a comprehensive understanding and manipulation of biofilm formation is crucial for various applications, including wastewater treatment~\cite{Ghosh2019}, bioremediation~\cite{Radwan2002}, and biomedical engineering~\cite{Khalid2020,Zhou2022}.

Understanding the dynamics of bacteria near surfaces is fundamental for various biological and engineering processes~\cite{Li2009,Petroff2015,Perez2019,Junot2022}. In recent decades, extensive experimental, numerical and analytical studies have significantly developed our understanding of bacterial dynamics near surfaces with various properties~\cite{Giacche2010,Spagnolie2012,Molaei2014,Eisenstecken2016,Qi2017,Petroff2018,Das2019,Park2019,Zhang2021,Leishangthem2024}. Experimental observations suggest that bacterial surface entrapment involves three main stages: approach, reorientation, and surface swimming~\cite{Bianchi2017,Bianchi2019,Mousavi2020}. However, hydrodynamic interactions alone are insufficient to explain the stable swimming behavior of bacteria near solid surfaces~\cite{Frymier1995,Lauga2006}. In addition to hydrodynamic interactions, van der Waals and electrostatic interactions, as described by the Derjaguin–Landau–Verwey–Overbeek (DLVO) theory, also play essential roles in bacterial behavior near surfaces~\cite{Frymier1995,Vigeant1997,Mcclaine2002,Vigeant2002,Li2008,Vissers2018,Shan2020}. Specifically, van der Waals forces dominate at short separations, while electrostatic forces become increasingly dominant at larger distances~\cite{Sharma2003,Hong2012}. 

Experiments report two distinct types of measured heights for bacteria swimming near solid surfaces. The first type is the height of the stable fixed point, reported values across different experiments typically range from a few nanometers to more than one hundred nanometers. For~\emph{E. coli}, reported stable heights generally range from a few nanometers up to tens of nanometers, and the reasons for this dispersion remain unclear~\cite{Frymier1995,Vigeant1997,Mcclaine2002,Lauga2006,Petroff2018}. The second type corresponds to instantaneous heights recorded while a bacterium performs approximately circular motion near the surface. These instantaneous heights are highly variable, spanning a few nanometers to several hundred nanometers, because near-field hydrodynamic interactions tend to confine bacteria near the surface while Brownian motion causes fluctuations around the stable fixed point. Additional contributions to this dispersion include surface entrapment, bacterial species differences, and environmental conditions~\cite{Lauga2006,Li2008,Bianchi2017,Liu2024,Leishangthem2024}.

Under the no‑slip boundary condition, the tangential fluid velocity at the interface is equal to that of the surface~\cite{Lauga2007}. As bacteria swim near these surfaces, shear flow exerts a torque on them, leading to the orientation of their cell bodies toward the surface, yet flagella hinder this rotation~\cite{Das2019,Bianchi2017,Sipos2015}. Bacteria reach a stable fixed point under the combined action of hydrodynamic forces acting on the cell body and flagellum, as well as electrostatic and van der Waals interactions described by DLVO theory. This point is specified by the height of the bacteria above the surface and the inclination angle of the flagellar axis relative to the surface~\cite{Li2008,Shum2010}. DLVO theory can be used to evaluate how the ionic strength affects the stable swimming behavior of bacteria on surfaces. The Debye length, a critical parameter in this theory, is strongly dependent on the ionic strength of the electrolyte solution~\cite{Sharma2003,Hong2012,Lavaud2021}. Moreover, the flagellar morphology of bacteria markedly influences their motility patterns on surfaces~\cite{Shum2010,Pimponi2016,Tokarova2021}, including the radius of curvature of the circular motion observed in experiments. These radii of curvature are typically in the range of $8-50$~\si{\mu m}~\cite{Frymier1995,Vigeant1997,Shum2010,Dunstan2012,Pimponi2016,Tokarova2021}. However, it is unclear how the flagellar morphology and the ionic strength of the electrolyte solution collaboratively affect the stable height, inclination angle, and radius of curvature of the circular motion.

Monotrichous bacteria, with a single helical flagellum, are particularly suitable for simulation modeling due to their simple morphology~\cite{Li2008,Shum2010,Shum2015,Vizsnyiczai2020}. To systematically explore the influence of the flagellar morphology and ionic strength of the electrolyte solution on stable circular motion, we develop a mathematical model of a bacterium with a rigid helical flagellum swimming near a solid surface. A detailed description of this model and the numerical simulation methods is provided in Sec.~\uppercase\expandafter{\romannumeral2. A}, and the DLVO theory is presented in Sec.~\uppercase\expandafter{\romannumeral2. B}. First, we examine the three main stages of the bacterial surface entrapment in Sec.~\uppercase\expandafter{\romannumeral3. A}. Subsequently, Secs.~\uppercase\expandafter{\romannumeral3. B} and \uppercase\expandafter{\romannumeral3. C} investigate the stable fixed point and the radius of curvature of the stable circular motion, respectively. Finally, the main conclusions are summarized in Sec.~\uppercase\expandafter{\romannumeral4}.

\section{Model and Methods}
\subsection{Bacterial Model}
\begin{figure}
\centering
\includegraphics[width=0.50\textwidth]{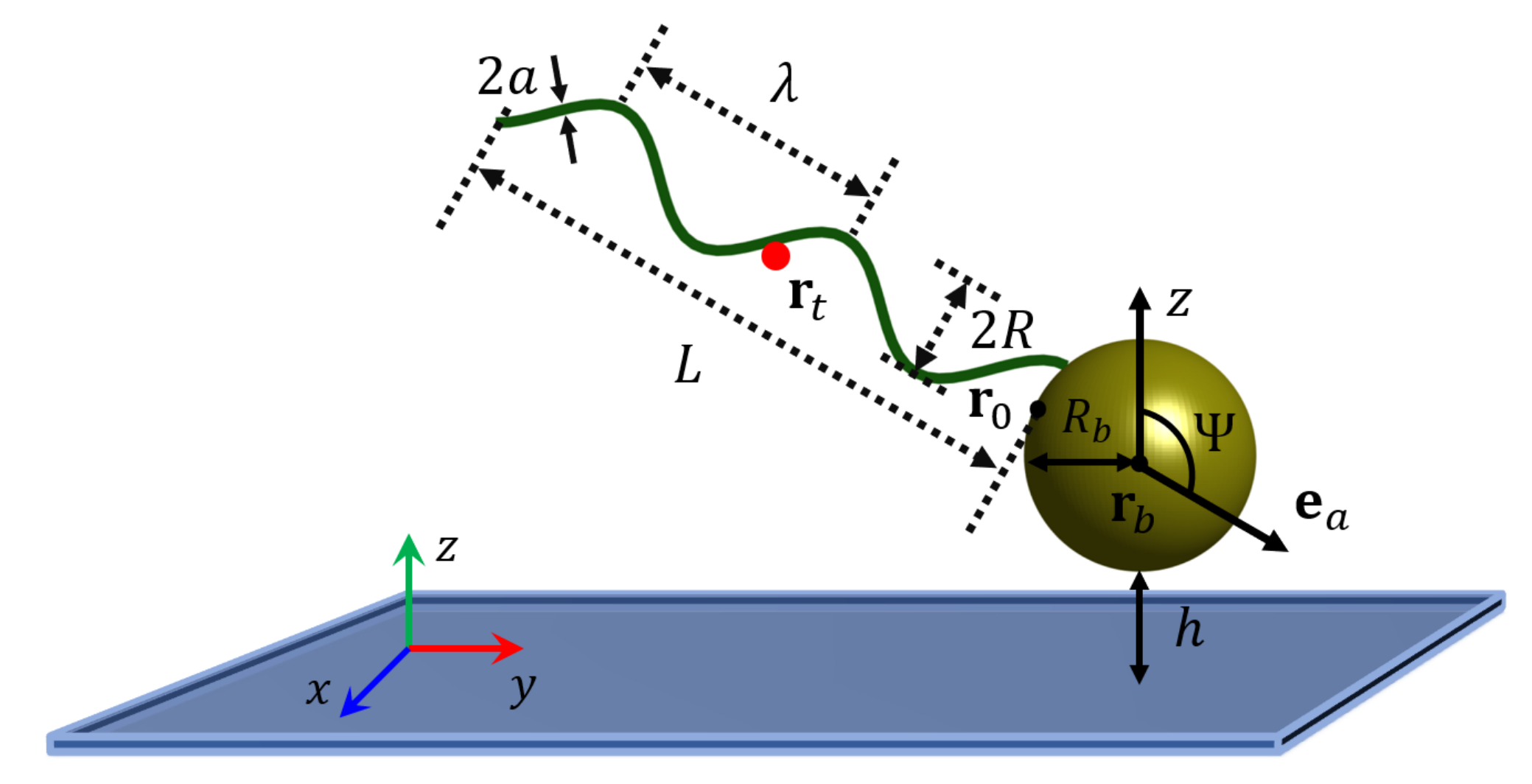}
\caption{Schematic of a bacterium near a solid surface. The height $h$ represents the minimum gap between the cell body and the solid surface. The inclination angle $\Psi$ is defined as the angle between the direction of the flagellar axis direction and the $z$-axis. The radius of the cell body is $R_{b}$. The helical flagellum, with a contour length $\Lambda$, helix radius $R$, pitch $\lambda$, pitch angle $\theta$, filament radius $a$, and axis length $L=\Lambda\cos\theta$, where $\tan\theta=2\pi R/\lambda$. The centers of the cell body and flagellum are $\mathbf{r}_{b}$ and $\mathbf{r}_{t}$, respectively.}
\label{fig:fig1}
\end{figure}

Bacteria swimming in a fluid medium are in the low Reynolds number regime (typically $Re\approx10^{-4}$), with their dynamics described by the incompressible Stokes equations~\cite{Lauga2006}. The viscous forces and torques for various parts of flagellated bacteria are linearly related to their translational and rotational velocities and are expressed as~\cite{Lauga2009,Dunstan2012,Liu2025A}:
\begin{equation}
\left(\begin{matrix}\mathbf{F}_{b}\\ \mathbf{T}_{b} \\ \mathbf{F}_{t}\\ \mathbf{T}_{t} \end{matrix}\right)=\mathcal{R}\left(\begin{matrix}\mathbf{U}_{b}\\ \mathbf{W}_{b}\\ \mathbf{U}_{t}\\ \mathbf{W}_{t} \end{matrix}\right),
\label{eq:refname01}
\end{equation}
where $\mathbf{U}_{b}$ and $\mathbf{W}_{b}$ are the instantaneous translational and rotational velocities of the cell body, while $\mathbf{U}_{t}$ and $\mathbf{W}_{t}$ represent those of the flagellum, respectively. $\mathbf{F}_{b}$, $\mathbf{T}_{b}$, $\mathbf{F}_{t}$, and $\mathbf{T}_{t}$ are the forces and torques exerted on the fluid by the cell body and the flagellum, respectively. In this study, the bacterium model consists of a spherical cell body and a rigid helical flagellum, as illustrated in Fig.~\ref{fig:fig1}. The connection point between the flagellar axis and the surface of the cell body is defined as $\mathbf{r}_{0}$. The radius of the cell body is $R_{b}$, and the centers of the cell body and flagellum are $\mathbf{r}_{b}$ and $\mathbf{r}_{t}$, respectively. The helical flagellum is modeled as a rigid helix with radius $R$, pitch $\lambda$, pitch angle $\theta$, filament radius $a$, axial length $L$, and contour length $\Lambda=L/\cos\theta$, where $\tan\theta=2\pi R/\lambda$. In the laboratory frame, the centerline positions of the left-handed helical flagellum are given by~\cite{Liu2025B}:
\begin{equation}
\begin{split}
\mathbf{r}(s)=-s\cos\theta\mathbf{D}_{1}+R\cos\phi\mathbf{D}_{2}+R\sin\phi\mathbf{D}_{3}+\mathbf{r}_{0},
\label{eq:refname02}
\end{split}
\end{equation}
where $\phi=ks\cos\theta$ is the phase and $k=2\pi/\lambda$ is the wavenumber. The parameter $s\in[0,\Lambda]$ is the arc length measured from the connection point $\mathbf{r}_{0}$, along the flagellar centerline. The associated material frame at the connection point $\mathbf{r}_{0}$ is defined as $\{\mathbf{D}_{1},\mathbf{D}_{2},\mathbf{D}_{3}\}$, where $\mathbf{D}_{1}$ is the unit vector along the flagellar axis pointing towards the cell body~\cite{Liu2025B}.

We treat the resistance matrices of the cell body and the flagellum separately, neglecting the hydrodynamic interactions between them~\cite{Lauga2006} and between the flagellum and the surface. The filament radius $a$ is about $10$~\si{nm}\cite{Darnton2007}, and the hydrodynamic interactions between the flagellum and a nearby surface become significant only when the separation is comparable to or smaller than the filament radius $a$~\cite{Dunstan2012}. Since such separations rarely fall below this value for non-colonizing bacteria~\cite{Bianchi2017,Khalid2020}, these interactions can be reasonably neglected. The resistance matrix, $\mathcal{R}$, of a bacterial system can be expressed as:
\begin{equation}
\begin{split}
\mathcal{R}=\left(\begin{matrix} 
\mathcal{R}_{b} & 0 \\
0 & \mathcal{R}_{t}
\end{matrix}\right).
\label{eq:refname03}
\end{split}
\end{equation}
The submatrices $\mathcal{R}_{b}$ and $\mathcal{R}_{t}$ are resistance matrices of the cell body and flagellum, respectively. As shown in Fig.~\ref{fig:fig1}, the resistance matrix, $\mathcal{R}_{b}$, of a spherical cell body near a solid surface is given by~\cite{Dunstan2012}:
\begin{equation}
\begin{split}
\mathcal{R}_{b}=\left(\begin{matrix} 
Y_{\parallel}^{A} & 0 & 0 & 0 & Y^{B} & 0 \\
0 & Y_{\parallel}^{A} & 0 & -Y^{B} & 0 & 0 \\
0 & 0 & Y_{\perp}^{A} & 0 & 0 & 0 \\
0 & -Y^{B} & 0 & Y_{\parallel}^{C} & 0 & 0 \\
Y^{B} & 0 & 0 & 0 & Y_{\parallel}^{C} & 0\\
0 & 0 & 0 & 0 & 0 & Y_{\perp}^{C} 
\end{matrix}\right).
\label{eq:refname04}
\end{split}
\end{equation}
The detailed expressions of $Y_{\parallel}^{A}$, $Y_{\perp}^{A}$, $Y^{B}$, $Y_{\parallel}^{C}$, and $Y_{\perp}^{C}$, which represent the translational and rotational resistance coefficients of the cell body, are presented in Appendix~\ref{appB}. Note that Eq.~\ref{eq:refname01}, which describes the motion of bacteria near a solid surface, is a system of nonlinear equations.

A helical flagellum is simplified into a chiral body, and its $6\times6$ resistance matrix, $\mathcal{R}_{t}$, is calculated using resistive force theory (RFT)~\cite{Gray1955,Chwang1975,Johnson1979}. The resistance matrix is expressed as~\cite{Di2011,Liu2026A}:
\begin{equation}
\begin{split}
\mathcal{R}_{t}=\left(\begin{matrix} 
A_{f} & B_{f}^{T} \\
B_{f} & C_{f}
\end{matrix}\right),
\label{eq:refname05}
\end{split}
\end{equation}
where the superscript "T" denotes matrix transpose. The submatrices of the flagellar resistance matrix are defined as
\begin{equation}
\begin{split}
&A_{f}=X_{\parallel}^{A}\mathbf{e}_{a}\otimes\mathbf{e}_{a}+X_{\perp}^{A}\left(\mathbb{I}-\mathbf{e}_{a}\otimes\mathbf{e}_{a}\right),\\
&B_{f}=X_{\parallel}^{B}\mathbf{e}_{a}\otimes\mathbf{e}_{a}+X_{\perp}^{B}\left(\mathbb{I}-\mathbf{e}_{a}\otimes\mathbf{e}_{a}\right),\\
&C_{f}=X_{\parallel}^{C}\mathbf{e}_{a}\otimes\mathbf{e}_{a}+X_{\perp}^{C}\left(\mathbb{I}-\mathbf{e}_{a}\otimes\mathbf{e}_{a}\right),
\label{eq:refname06}
\end{split}
\end{equation}
where $\mathbf{e}_{a}$ represents the unit directional vector of the flagellar axis, with the subscript "f" indicating the flagellum. Detailed expressions for $X_{\parallel}^{A}$, $X_{\perp}^{A}$, $X_{\parallel}^{B}$, $X_{\perp}^{B}$, $X_{\parallel}^{C}$, and $X_{\perp}^{C}$ are provided in Appendix~\ref{appC}. 

The instantaneous translational and rotational velocities of the cell body are defined as $\mathbf{U}_{b}=(U_{x},U_{y},U_{z})$ and $\mathbf{W}_{b}=(W_{x},W_{y},W_{z})$, respectively. The corresponding velocities of the flagellar center are given by:
\begin{equation}
\begin{split}
&\mathbf{W}_{t}=\mathbf{W}_{b}+\mathbf{W}_{0},\\
&\mathbf{U}_{t}=\mathbf{U}_{b}+\mathbf{W}_{t}\times(\mathbf{r}_{t}-\mathbf{r}_{b}),
\label{eq:refname07}
\end{split}
\end{equation}
where $\mathbf{W}_{0}=2\pi f\mathbf{e}_{m}$ represents the motor rotation velocity. $\mathbf{r}_{b}$ and $\mathbf{r}_{t}$ denote the positions of the centers of the cell body and the flagellum, respectively, with the distance between them defined as $l=L/2+R_{b}$. The bacterium is modeled as a chiral two-body model~\cite{Di2011,Liu2026A}. The force and torque balance equations for a bacterium swimming near a solid surface are given by:
\begin{equation}
\begin{split}
&\mathbf{F}_{b}+\mathbf{F}_{t}-\mathbf{F}_{b}^{ext}-\mathbf{F}_{t}^{ext}=0,\\
&\mathbf{T}_{b}+\mathbf{T}_{t}-\mathbf{T}_{b}^{ext}-\mathbf{T}_{t}^{ext}+\left(\mathbf{r}_{t}-\mathbf{r}_{b}\right)\times\left(\mathbf{F}_{t}-\mathbf{F}_{t}^{ext}\right)=0,
\label{eq:refname08}
\end{split}
\end{equation}
where $\mathbf{F}_{b}^{ext}$, $\mathbf{T}_{b}^{ext}$, $\mathbf{F}_{t}^{ext}$, and $\mathbf{T}_{t}^{ext}$ are external forces and torques exerted on the cell body and flagellum, respectively, excluding hydrodynamic forces and torques. The morphological parameters of the bacteria and their corresponding values used in our simulations are summarized in Table~\ref{tab:table1}.

\begin{table}
\caption{\label{tab:table1}Symbols, definitions and corresponding values in this article.}
\begin{ruledtabular}
\begin{tabular}{cll}
\textbf{Notation} & \textbf{Description} & \textbf{Value}\\
\hline
$\mu$ & Dynamic viscosity & $1.0$ \si{\mu g/(\mu m\cdot s)}\\
$R_{b}$ & Radius of cell body & $1.0$ \si{\mu m}\\
$R$ & Helix radius & $0.25$ \si{\mu m}\\
$a$ & Filament radius & $0.01$ \si{\mu m}\\
$\theta$ & Pitch angle & $\pi/5$\\
$f$ & Motor rotation frequency & $100$ \si{Hz}\\
$\mathbf{e}_{a}$ & Flagellar axis direction & $y$-axis\\
$\mathbf{e}_{m}$ & Rotation direction of motor & $\mathbf{e}_{m}=\frac{\mathbf{r}_{0}-\mathbf{r}_{b}}{|\mathbf{r}_{0}-\mathbf{r}_{b}|}$\\
$h$ & Height above the surface & \\
$\Psi$ & inclination angle & $0\sim\pi$\\
\end{tabular}
\end{ruledtabular}
\end{table}

\subsection{DLVO Theory}
The interactions between the cell body and the surface include hydrodynamic forces, van der Waals forces, and electrostatic forces when bacteria swim near a solid surface. The latter two interactions are described by DLVO theory~\cite{Mcclaine2002,Vigeant2002,Li2008,Khalid2020,Krsmanovic2021}. The DLVO force, $F_{D}$, between a spherical cell body and a solid surface is the sum of the attractive van der Waals (vdW) force and the repulsive or attractive electrostatic (ele) force:
\begin{equation}
\begin{split}
&F_{D}(h)=F_{vdW}(h)+F_{ele}(h),
\label{eq:refname09}
\end{split}
\end{equation}
where $h$ is the minimum gap between the cell body and the solid surface. The subscripts indicate different types of interactions. The mathematical expressions for these interaction energies are given as follows~\cite{Sharma2003,Hong2012}:
\begin{equation}
\begin{split}
&G_{vdW}(h)=-\frac{H}{12}\left[\frac{2R_{b}(h+R_{b})}{h(h+2R_{b})}+\ln\left(\frac{h}{h+2R_{b}}\right)\right],\\
&G_{ele}(h)=\pi\varepsilon R_{b}\left[2\zeta_{1}\zeta_{2}\ln\left(\frac{1+e^{-\kappa h}}{1-e^{-\kappa h}} \right)+\left(\zeta_{1}^{2}+\zeta_{2}^{2}\right)\ln\left(1-e^{-2\kappa h}\right)\right].
\label{eq:refname10}
\end{split}
\end{equation}
The corresponding forces are obtained by differentiating with respect to $h$:
\begin{equation}
\begin{split}
&F_{vdW}(h)\equiv-\frac{\partial G_{vdW}(h)}{\partial h}=-\frac{H}{3}\frac{R_{b}^{3}}{h^{2}(h+2R_{b})^{2}},\\
&F_{ele}(h)\equiv-\frac{\partial G_{ele}(h)}{\partial h}=2\pi\varepsilon\kappa R_{b}\frac{2\zeta_{1}\zeta_{2}e^{\kappa h}-(\zeta_{1}^{2}+\zeta_{2}^{2})}{e^{2\kappa h}-1},
\label{eq:refname11}
\end{split}
\end{equation}
where $R_{b}$ is the radius of the cell body, $H$ is the Hamaker constant, and $\zeta_{1}$ and $\zeta_{2}$ are the zeta potentials of the cell body and the solid surface, respectively. Typical zeta potential values are provided in Table~\ref{tab:table2}. $\kappa$ denotes the inverse Debye length, which is given by~\cite{Hong2012,Lavaud2021}
\begin{equation}
\begin{split}
&\kappa^{-1}=\sqrt{\frac{\varepsilon k_{B}T}{2e^{2}N_{A}I}}\approx\frac{0.304}{\sqrt{I}} \si{nm},
\label{eq:refname12}
\end{split}
\end{equation}
where $\varepsilon=6.933\times10^{-10}$~\si{C^{2}/(N\cdot m^{2})} is the permittivity of water at room temperature $T=298$~\si{K}, $k_{B}=1.381\times10^{-23}$~\si{J/K} is the Boltzmann constant, $e=1.602\times10^{-19}$~\si{C} is the elementary charge, and $N_{A}=6.022\times10^{23}$~\si{mol^{-1}} is the Avogadro constant. $I$ is the ionic strength in~\si{mol/L} (or~\si{M}). The values of ionic strength and zeta potential are listed in Table~\ref{tab:table2}. Eq.~\ref{eq:refname12} indicates that lower ionic strengths correspond to larger Debye lengths.

\begin{figure}
\centering
\includegraphics[width=0.50\textwidth]{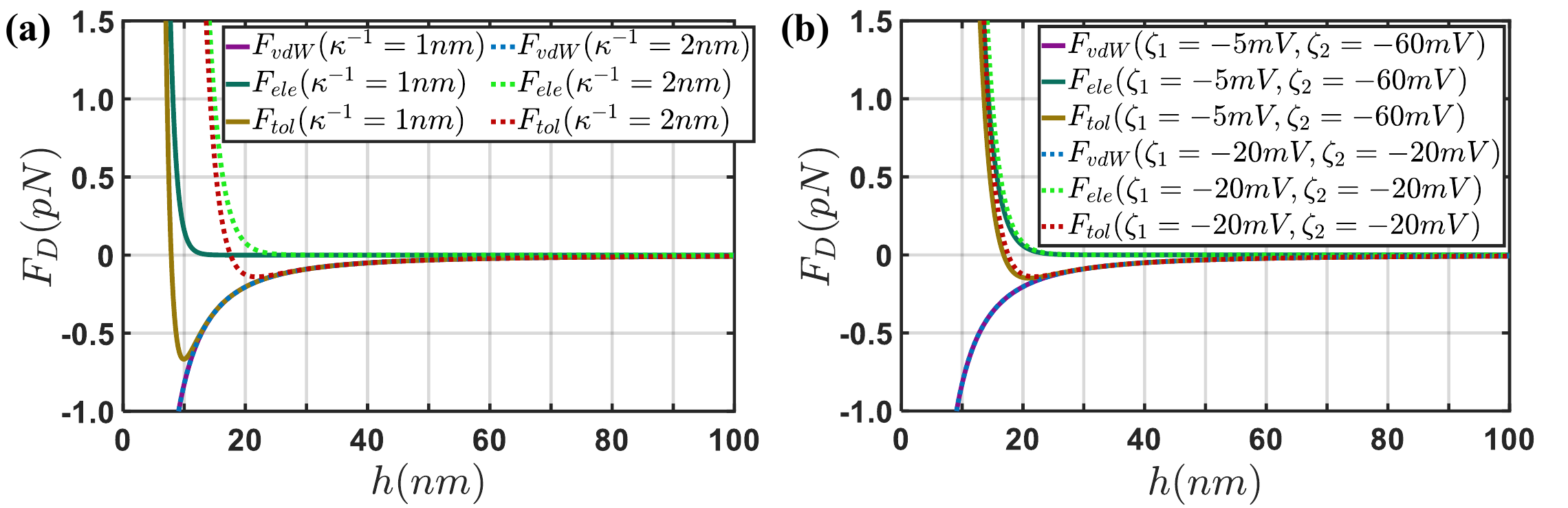}
\caption{DLVO force as a function of the separation distance $h$ between a spherical cell body and a solid surface. (a) DLVO force for different Debye lengths, with $\zeta_{1}=-20$~\si{mV} and $\zeta_{2}=-20$~\si{mV}. (b) DLVO force for different zeta potentials, with $\kappa^{-1}=2$~\si{nm}.}
\label{fig:fig2}
\end{figure}

\begin{table}
\caption{\label{tab:table2}Typical values of Debye length corresponding to different ionic strength $I$ at $298$~\si{K}. Zeta potentials for various bacteria and surface.}
\begin{ruledtabular}
\begin{tabular}{cccc}
$I$ (\si{mol/L}) & $\kappa^{-1}$ (\si{nm}) & $\zeta_{1}$ (Bacterium) (\si{mV}) & $\zeta_{2}$ (Surface) (\si{mV})\\
\hline
$0.006$ & $3.92$  & $-22.82$  & $-37.29$ ~\cite{Vigeant1997}\\
$0.006$ & $3.92$  & $-25.1$   & $-56.8$  ~\cite{Vigeant2002}\\
$0.01$  & $3.04$  & $-30\pm1$ & $-21\pm2$~\cite{Shan2020}\\
$0.01$  & $3.04$  & $-53\pm2$ & $-21\pm2$~\cite{Shan2020}\\
$0.024$ & $2.00$  & $-20.0$   & $-20.0$  ~\cite{Li2008}\\
$0.05$  & $1.36$  & $-14\pm2$ & $-12\pm1$~\cite{Shan2020}\\
$0.05$  & $1.36$  & $-43\pm2$ & $-12\pm1$~\cite{Shan2020}\\
$0.1$   & $0.96$  & $-11\pm1$ & $-8\pm1$ ~\cite{Shan2020}\\
$0.1$   & $0.96$  & $-36\pm3$ & $-8\pm1$ ~\cite{Shan2020}\\
$0.19$  & $0.697$ & $-21.5$   & $-36.2$  ~\cite{Frymier1995}\\
$0.20$  & $0.680$ & $-19.6$   & $-18.7$  ~\cite{Vigeant2002}\\
$0.202$ & $0.676$ & $-4.77$   & $-14.29$ ~\cite{Vigeant1997}\\
\end{tabular}
\end{ruledtabular}
\end{table}

The Hamaker constant $H$ ranges from $10^{-21}$~\si{J} to $10^{-19}$~\si{J}. In this work, we set $H=10^{-21}$~\si{J}, consistent with previous studies~\cite{Frymier1995,Li2008}. Fig.~\ref{fig:fig2} presents the DLVO force profiles for different Debye lengths $\kappa^{-1}$ and zeta potentials $\zeta_{1}$ and $\zeta_{2}$. The left panel of Fig.~\ref{fig:fig2} presents the van der Waals and electrostatic forces for $\zeta_{1}=\zeta_{2}=-20$~\si{mV}, with Debye lengths of $\kappa^{-1}=1$~\si{nm} and $\kappa^{-1}=2$~\si{nm}. This panel reveals that the DLVO forces are sensitive to the Debye length. The right panel shows the forces for different zeta potentials, with a fixed Debye length of $\kappa^{-1}=2$~\si{nm}. As presented in Table~\ref{tab:table2}, the measured zeta potentials vary among experiments. Furthermore, we find that the DLVO forces show less sensitivity to the zeta potentials of the cell body and solid surface compared to their sensitivity to the Debye length, as illustrated in Fig.~\ref{fig:fig2}(b) and Eq.~\ref{eq:refname11}. For simplicity, we set $\zeta_{1}=\zeta_{2}=-20$~\si{mV} and focus on the effects of Debye length on bacterial motility near a solid surface~\cite{Li2008}.

\section{Results and Discussion}
\subsection{Three Stages of Surface Entrapment}
The trajectories of bacteria swimming near a solid surface are determined by their initial positions and orientations. Because the space of initial conditions is continuous and effectively infinite, it is impractical to perform long simulations for all possible initial conditions to explore the complete picture. The height $h$ and the inclination angle $\Psi$ are used to simplify the analysis of bacterial dynamics near a solid surface. In addition, we perform phase averaging over one flagellar rotation period to eliminate the rapid oscillations associated with the flagellar phase, which varies on a much faster time scale than translational motion~\cite{Shum2010,Pimponi2016}. Monotrichous bacteria can be simplified into a chiral two-body model, which is obtained by integrating the centerline of the flagellum using RFT~\cite{Di2011,Dvoriashyna2021,Liu2026A}. Neglecting Brownian motion, bacterial trajectories near the surface are simulated using Eq.~\ref{eq:refname01}. Then the center position of the cell body is updated iteratively as follows~\cite{Liu2025B}:
\begin{equation}
\begin{split}
&\mathbf{r}_{b}(t+\Delta t)=\mathbf{r}_{b}(t)+\mathbf{U}_{b}(t)\Delta t.
\label{eq:refname13}
\end{split}
\end{equation}
The direction of the flagellar axis, $\mathbf{e}_{a}$, is updated by rotation:
\begin{equation}
\begin{split}
&\mathbf{e}_{a}(t+\Delta t)=\mathbf{R}(\mathbf{e},\Delta\varphi)\cdot\mathbf{e}_{a}(t),
\label{eq:refname14}
\end{split}
\end{equation}
where $\mathbf{e}=\mathbf{W}_{b}/|\mathbf{W}_{b}|$ is the unit vector along the rotation direction and $\Delta\varphi=|\mathbf{W}_{b}|\Delta t$ is the rotation angle of the cell body. The rotation matrix $\mathbf{R}(\mathbf{e},\Delta\varphi)$ is Rodrigues' rotation matrix, defined as~\cite{Dai2015,Murray2017,Liu2025B}
\begin{equation}
\begin{split}
&\mathbf{R}(\mathbf{e},\Delta\varphi)=\cos\Delta\varphi\mathbb{I}+(1-\cos\Delta\varphi)\mathbf{e}\otimes\mathbf{e}^{T}+\sin\Delta\varphi\mathbf{e}\times,
\label{eq:refname15}
\end{split}
\end{equation}
where $\mathbb{I}$ is a $3\times3$ identity matrix, $\otimes$ is Kronecker product, and $(\mathbf{e}\times)$ is a $3\times3$ antisymmetric matrix defined such that $(\mathbf{e}\times)\mathbf{v}=\mathbf{e}\times\mathbf{v}$ for any vector $\mathbf{v}$. The position of the flagellar center is updated by:
\begin{equation}
\begin{split}
&\mathbf{r}_{t}(t+\Delta t)=\mathbf{r}_{b}(t+\Delta t)-l\mathbf{e}_{a}(t+\Delta t).
\label{eq:refname16}
\end{split}
\end{equation}
The motor rotation direction is opposite to the flagellar axis, that is, $\mathbf{e}_{m}=-\mathbf{e}_{a}$.

\begin{figure}
\centering
\includegraphics[width=0.50\textwidth]{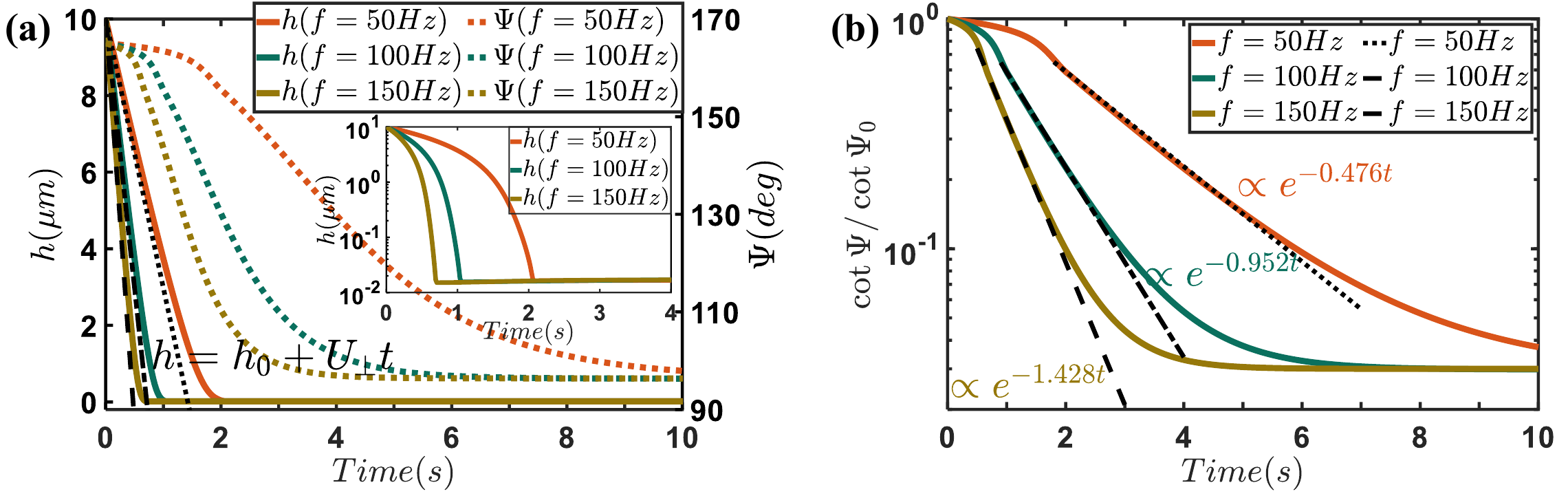}
\caption{(a) Time evolution of the height $h$ and inclination angle $\Psi$ for different motor rotation frequencies. The inset shows the time evolution of height $h$. The dashed black lines represent the analytical solutions. (b) Time evolution of normalized cotangent of inclination angle $\cot\Psi/\cot\Psi_{0}$. The dashed black lines show the fitted results.}
\label{fig:fig3}
\end{figure}

\begin{figure*}
\centering
\includegraphics[width=1.00\textwidth]{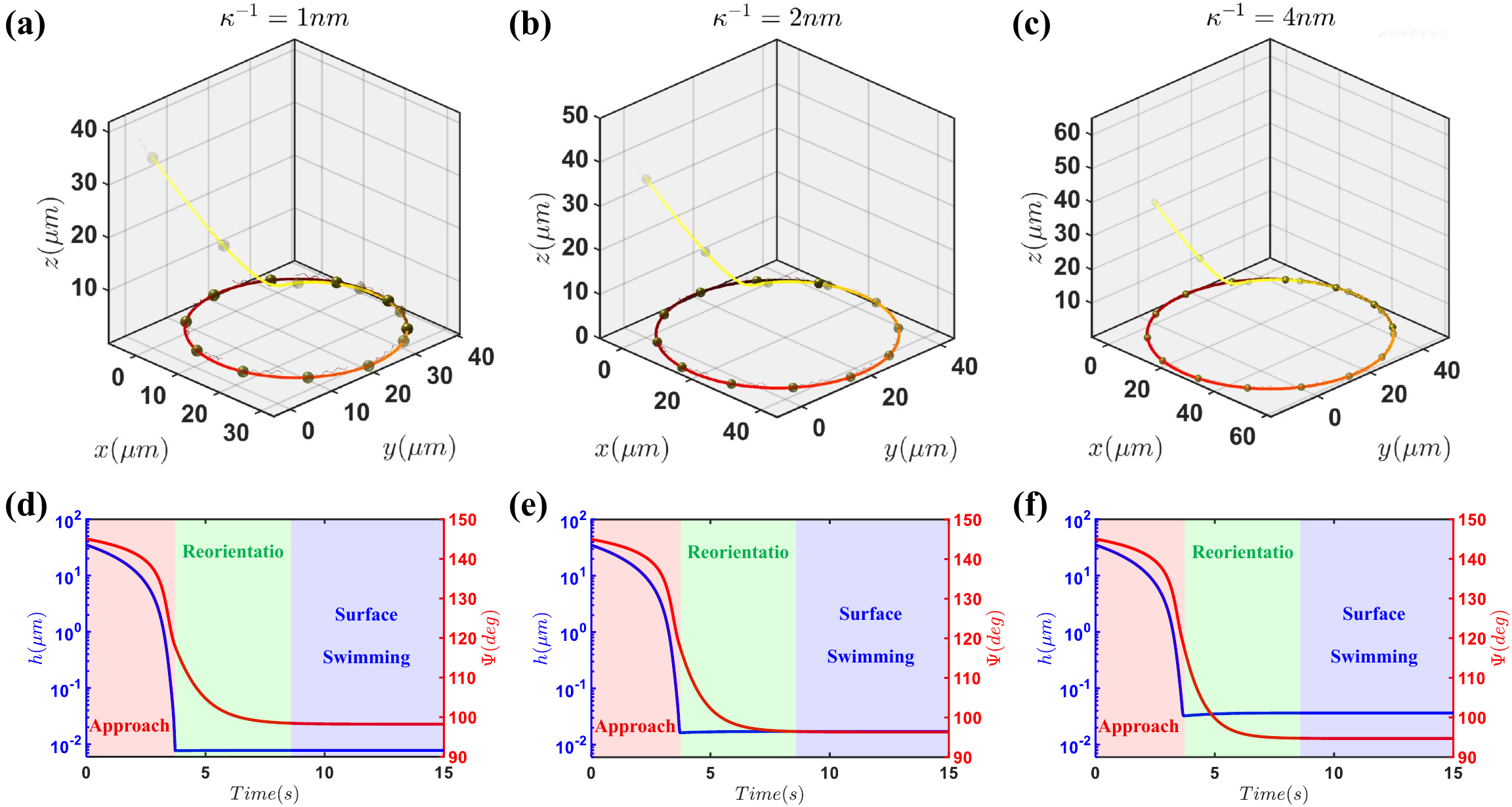}
\caption{Trajectories of a bacterium approaching a solid surface and exhibiting clockwise circular trajectories for different Debye lengths: (a) $\kappa^{-1}=1$~\si{nm}, (b) $\kappa^{-1}=2$~\si{nm}, and (c) $\kappa^{-1}=4$~\si{nm}, all starting from the same initial condition. The colour of the trajectory lines gradually darkens over time. (d-f) The corresponding heights $h$ (blue lines) and inclination angles $\Psi$ (red lines), both as functions of time.}
\label{fig:fig4}
\end{figure*}

In our simulations, the zeta potentials are $\zeta_{1}=\zeta_{2}=-20$~\si{mV} and the Debye length is $\kappa^{-1}=2$~\si{nm}. The bacterium is initially located at a height of $h_{0}=10$~\si{\mu m}, with an initial inclination angle of $\Psi_{0}=165^{\circ}$. The time step is $\Delta t=5\times10^{-4}$~\si{s}. Fig.~\ref{fig:fig3} (a) shows the time evolution of height $h$ and inclination angle $\Psi$ for various motor rotation frequencies. The morphological parameters of the bacteria are given in Table~\ref{tab:table1}, and the contour length of the flagellum is $\Lambda=7.5$~\si{\mu m}. This process can be divided into three main stages: approach, reorientation, and surface swimming. The translational velocities parallel ($\parallel$) and perpendicular ($\perp$) to the $xy$-plane in the bulk fluid are given by~\cite{Liu2026A}
\begin{equation}
\begin{split}
&U_{\parallel}=\frac{8\pi\mu R_{b}^{3}X_{\parallel}^{B}\cdot\sin\Psi}{(X_{\parallel}^{B})^{2}-(X_{\parallel}^{A}+6\pi\mu R_{b})(X_{\parallel}^{C}+8\pi\mu R_{b}^{3})}W_{0},\\
&U_{\perp}=\frac{8\pi\mu R_{b}^{3}X_{\parallel}^{B}\cdot\cos\Psi}{(X_{\parallel}^{B})^{2}-(X_{\parallel}^{A}+6\pi\mu R_{b})(X_{\parallel}^{C}+8\pi\mu R_{b}^{3})}W_{0},
\label{eq:refname17}
\end{split}
\end{equation}
where $W_{0}=-2\pi f$ is the motor rotation rate, implying that the flagellum rotates anticlockwise (viewed from behind the bacterium). The flagellar thrust of a swimming \emph{E. coli} is about $0.3$~\si{pN} obtained by the chiral two-body model. The black dashed lines in Fig.~\ref{fig:fig3}(a) represent the analytical solution $h=h_{0}+U_{\perp}t$, indicating that during the first stage bacteria exhibit quasi-linear motion with an approach velocity approximately proportional to the motor rotation frequency.

The motor rotation frequencies are set to $f=50$, $100$, and $150$~\si{Hz}. Here $\Psi_{0}$ denotes the initial inclination angle. During the reorientation stage, the height $h$ remains approximately constant, while the inclination angle $\Psi$ evolves over time, as shown in Fig.~\ref{fig:fig3}(a). Similarly, the time evolution of $\cot\Psi$ is illustrated in Fig.~\ref{fig:fig3}(b), the exponent of the exponential function is proportional to the product of the motor rotation frequency $f$ and time $t$. The relationship between the inclination angle $\Psi$, the motor rotation frequency $f$ and the time $t$ can be approximated by $\cot\Psi/\cot\Psi_{0}\propto e^{-\alpha f t}$, where $\alpha$ is approximately constant depending on the bacterial morphology and the height $h$~\cite{Mousavi2020}.

When swimming parallel to the solid surface, the bacteria maintain a stable height $h^{*}$ and inclination angle $\Psi^{*}$. To illustrate the effect of the ionic strength on the radius of curvature of stable circular motion, we perform numerical simulations using several commonly used Debye lengths: $\kappa^{-1}=1$~\si{nm}, $\kappa^{-1}=2$~\si{nm} and $\kappa^{-1}=4$~\si{nm}. The simulations shown in Fig.~\ref{fig:fig4} use a motor rotation frequency of $f=100$~\si{Hz}. The initial conditions are $h_{0}=35$~\si{\mu m} and $\Psi_{0}=145^{\circ}$.

Figs.~\ref{fig:fig4}(a-c) show the trajectories of a bacterium for different Debye lengths, highlighting the clockwise circular motion observed in the third stage. Figs.~\ref{fig:fig4}(d-f) depict the time evolution of height and inclination angle, illustrating the three main stages of bacterial motility. The radii of curvature of the stable circular trajectories are $r\approx19$~\si{\mu m}, $r\approx25$~\si{\mu m}, and $r\approx33$~\si{\mu m} for the Debye lengths $\kappa^{-1}=1$~\si{nm}, $\kappa^{-1}=2$~\si{nm}, and $\kappa^{-1}=4$~\si{nm}, respectively. The corresponding stable heights are $h^{*}=7.8$, $17.2$, and $36.4$~\si{nm}, and the stable inclination angles are $\Psi^{*}=98.2^{\circ}$, $96.4^{\circ}$, and $94.7^{\circ}$. These results indicate that as ionic strength decreases, the stable height and the radius of curvature increase, while the inclination angle decreases.

\subsection{Stable Fixed Point}
The kinematic characteristics of a bacterium swimming near a solid surface are completely described by the height $h$ and the inclination angle $\Psi$, which together define the phase plane $(h,\Psi)$. In this section, we constrain the direction of the flagellar axis, $\mathbf{e}_{a}$, to lie in the $yz$-plane, with the $z$-axis normal to the surface. Previous studies have demonstrated the existence of a stable fixed point $(h^{*},\Psi^{*})$, which satisfies the following equations~\cite{Shum2010,Shum2015}:
\begin{equation}
\begin{cases}
\dot{h}(h^{*},\Psi^{*})=U_{z}=0,\\
\dot{\Psi}(h^{*},\Psi^{*})=W_{x}=0.
\label{eq:refname18}
\end{cases}
\end{equation}

\begin{figure}
\centering
\includegraphics[width=0.50\textwidth]{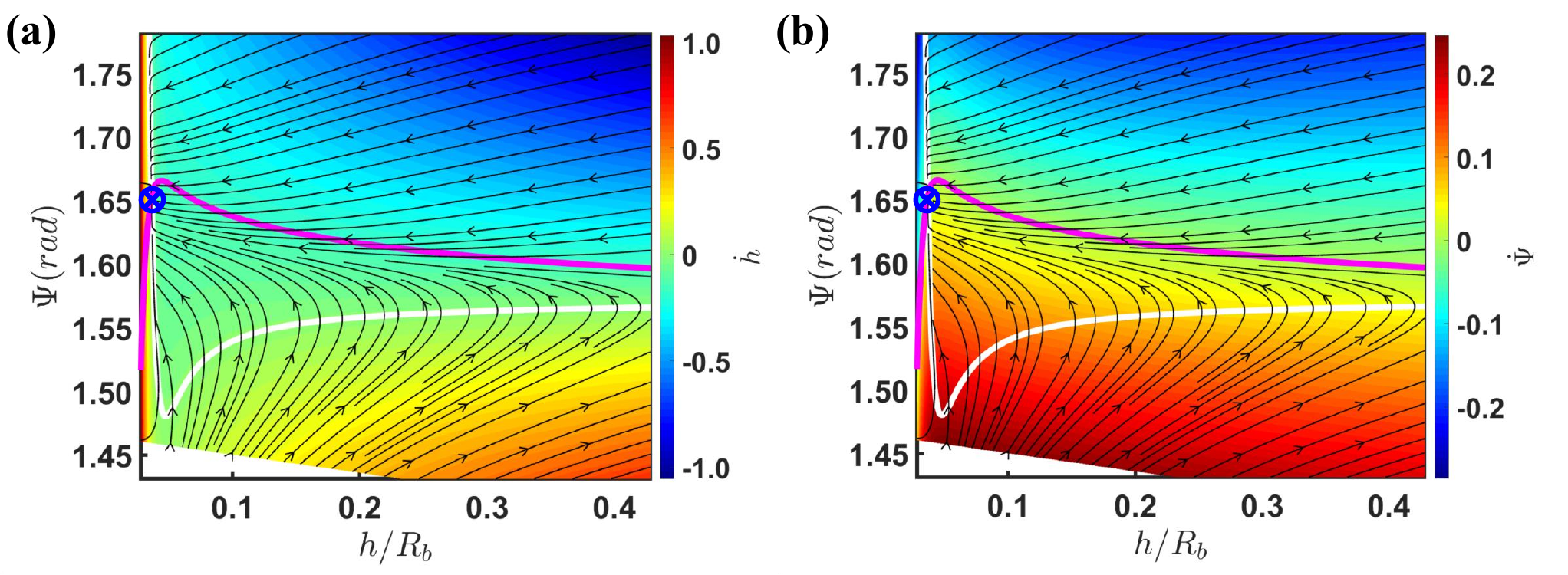}
\caption{Phase plane diagrams for a bacterium near a solid surface. The background colour maps represent (a) the translational velocity $\dot{h}$ and (b) the angular velocity $\dot{\Psi}$, respectively. The white and pink curves denote $\dot{h}=0$ and $\dot{\Psi}=0$, respectively, and their intersection point, indicated by symbol $\textcolor{blue}{\otimes}$, marks the stable fixed point}.
\label{fig:fig5}
\end{figure}

The phase plane for a bacterium swimming near a solid surface is presented in Fig.~\ref{fig:fig5}. In these simulations, the contour length of the flagellum is $\Lambda=7.5$~\si{\mu m}, and the Debye length is $\kappa^{-1}=4.0$~\si{nm}. The background color maps in Figs.~\ref{fig:fig5}(a) and (b) represent the values of $\dot{h}$ and $\dot{\Psi}$, respectively. The white and pink curves show the contours where $\dot{h}=0$ and $\dot{\Psi}=0$, respectively. The blank regions indicate areas that are inaccessible to the bacterium due to the presence of the solid surface. The figure shows a single stable fixed point $(h^{*},\Psi^{*})$ in the phase plane, with no other fixed points observed. Therefore, once the bacterium reaches the stable fixed point, its height $h$ and inclination angle $\Psi$ remain constant.

\begin{figure}
\centering
\includegraphics[width=0.50\textwidth]{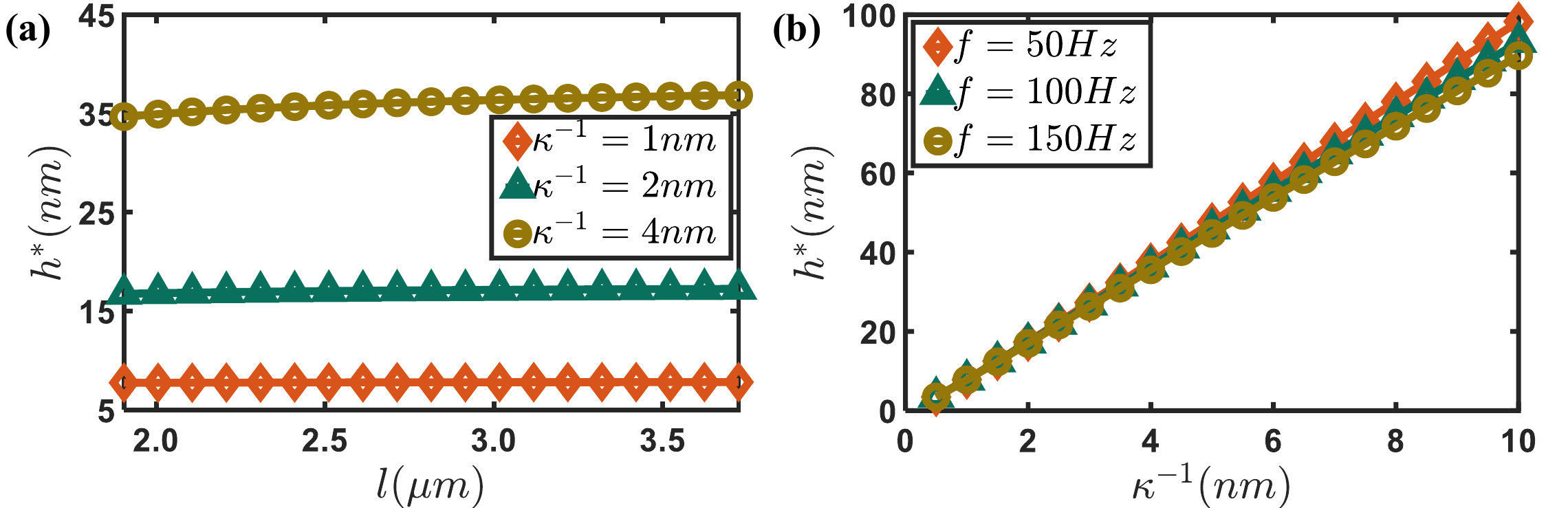}
\caption{The stable heights $h^{*}$ as functions of (a) the distance $l$ and (b) the Debye length $\kappa^{-1}$, respectively.}
\label{fig:fig6}
\end{figure}

As shown in Fig.~\ref{fig:fig4}, ionic strength significantly influences the stable height, the inclination angle, and the radius of curvature of stable circular motion near a solid surface. Experimental observations suggest that the radius of curvature is closely related to the stable height $h^{*}$~\cite{Lauga2006}. In this section, we investigate how flagellar morphology and ionic strength affect stable height $h^{*}$. Fig.~\ref{fig:fig6}(a) shows that $h^{*}$ depends only weakly on the flagellar contour length $\Lambda$, indicating that the effect of $\Lambda$ on $h^{*}$ is negligible over the parameter range considered.

Fig.~\ref{fig:fig6}(b) shows that the stable height $h^{*}$ is highly sensitive to the Debye length $\kappa^{-1}$, increasing with $\kappa^{-1}$. In our simulations, the contour length of the flagellum is $\Lambda=7.5$~\si{\mu m}. Under high ionic strength conditions ($I=0.1-0.3$~\si{M})\cite{Redman2004,Shan2020}, $h^{*}$ can be as small as about $10$~\si{nm}, whereas under low ionic strength conditions ($<0.05$~\si{M})\cite{Vigeant1997,Vigeant2002,Shan2020,Yee2000}, $h^{*}$ can reach several tens of nanometers. In general, the data in Fig.~\ref{fig:fig6}(b) indicate that $h^{*}$ increases approximately linearly with the Debye length $\kappa^{-1}$ over the range studied.

\begin{figure}
\centering
\includegraphics[width=0.50\textwidth]{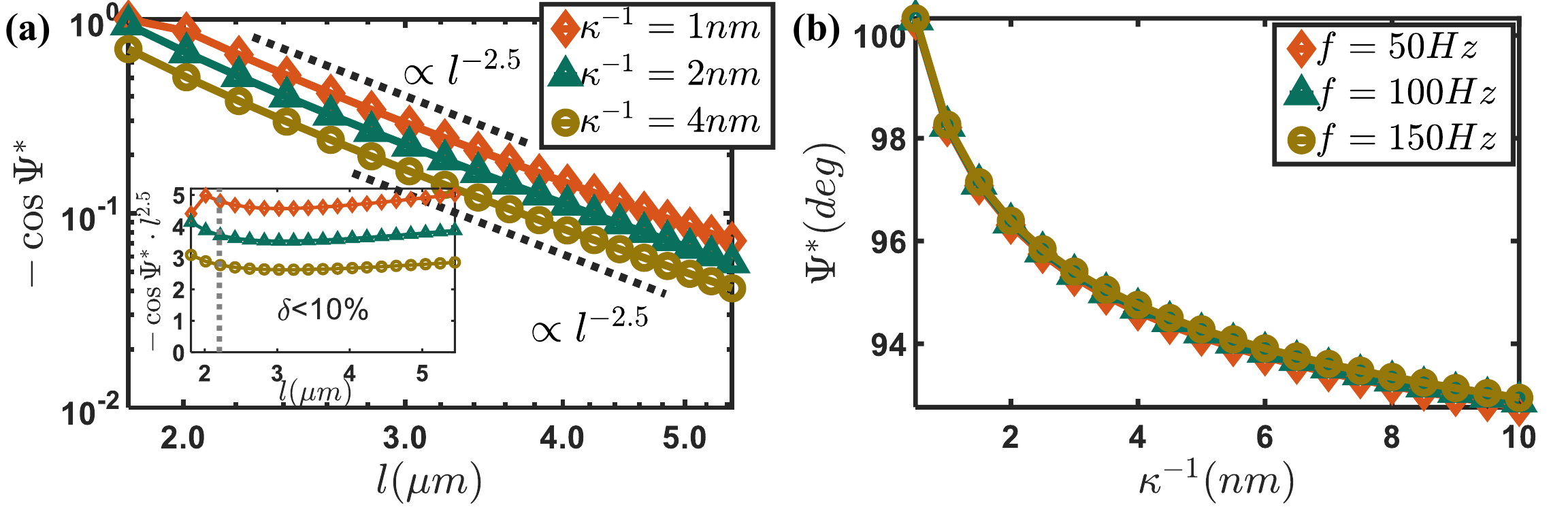}
\caption{(a) The negative cosine of the stable inclination angle, $-\cos\Psi^{*}$, as a function of the distance $l$. (b) The stable inclination angle, $\Psi^{*}$, as a function of Debye length $\kappa^{-1}$.}
\label{fig:fig7}
\end{figure}

In this section, we examine the relationship between the stable inclination angle $\Psi^{*}$ and the distance $l=L/2+R_{b}=\Lambda\cos\theta/2+R_{b}$ as well as the Debye length $\kappa^{-1}$. As illustrated in Fig.~\ref{fig:fig7}, the stable inclination angle $\Psi^{*}$ decreases monotonically with increasing $l$. For sufficiently short flagella, bacteria tend to point vertically toward the surface ("nose down"), consistent with experimental observations~\cite{Petroff2015}. This occurs because shorter flagella produce weaker torques that are insufficient to counteract rotation of the cell body~\cite{Petroff2018,Das2019}. The stable inclination angle $\Psi^{*}\approx96^{\circ}$ for $\Lambda=7.5$~\si{\mu m}. We define the relative deviation $\delta=[(\cos\Psi^{*}\cdot l^{2.5})_{max}-(\cos\Psi^{*}\cdot l^{2.5})_{min}]/(\cos\Psi^{*}\cdot l^{2.5})_{min}$. As shown in the inset of Fig.~\ref{fig:fig7}(a), $\delta<10\%$ for $l>2.2$~\si{\mu m}, indicating that $\cos\Psi^{*}$ approximately follows the empirical power law $-\cos\Psi^{*}\propto l^{-2.5}$ over the range studied. Fig.~\ref{fig:fig7}(b) shows the dependence of stable inclination angle $\Psi^{*}$ on the Debye length $\kappa^{-1}$ for a flagellar contour length of $\Lambda=7.5$~\si{\mu m}. As Debye length increases, the stable inclination angle decreases. 

When a bacterium swims near a solid surface, near-field hydrodynamic interactions generate a torque on the cell body that tends to rotate it toward the surface, while viscous resistance on the flagellum opposes that rotation. Because both torques scale approximately proportional to the motor frequency (as shown in Eq.~\ref{eq:refname17}), changing the motor frequency has little effect on the stable inclination angle determined by the torque balance. The vertical component of the flagellar propulsion force balances the DLVO repulsive force and thus determines the stable height. When the DLVO interaction is repulsive, it is highly sensitive to height, so small height changes produce large variations in repulsion, and variations in motor frequency therefore have only a weak effect on the stable height. As shown in Figs.~\ref{fig:fig6}(b) and ~\ref{fig:fig7}(b), the motor frequency has a negligible effect on the stable fixed point.

\subsection{Radius of Curvature of the Circular Trajectories}
\begin{figure}
\centering
\includegraphics[width=0.50\textwidth]{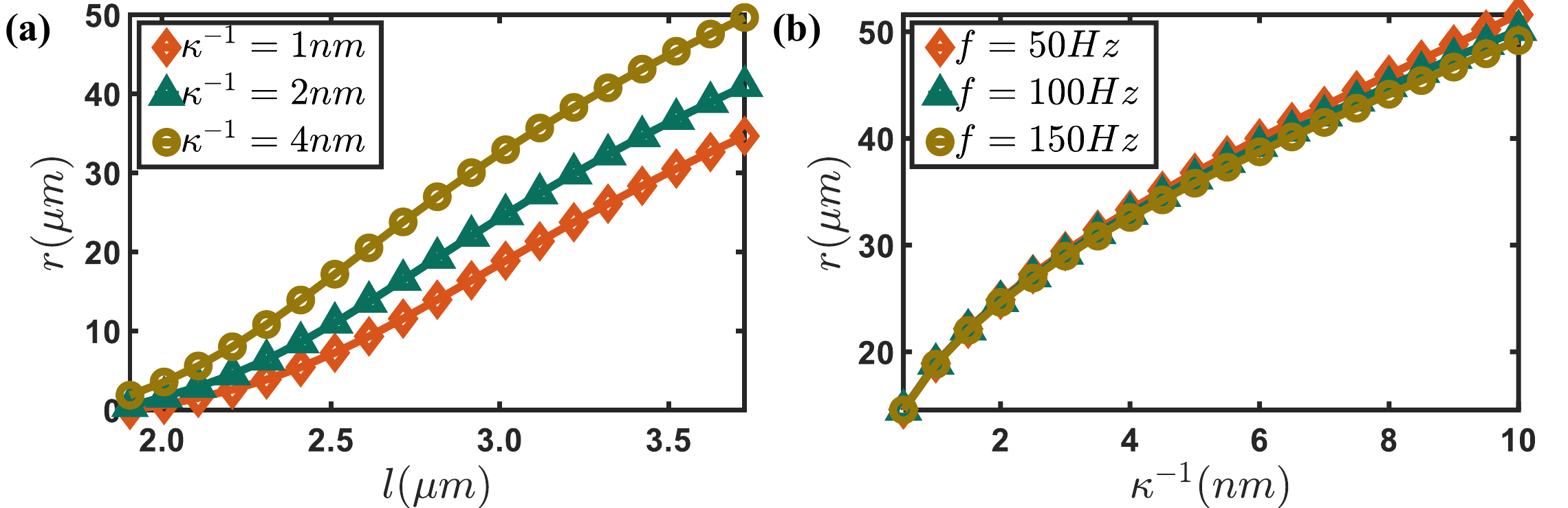}
\caption{The radii of curvature of the circular trajectories as functions of (a) the distance $l$ and (b) the Debye length $\kappa^{-1}$.}
\label{fig:fig8}
\end{figure}

When a bacterium swims near a solid surface, the direction of its circular motion is determined by the chirality of its flagellum. Specifically, if the flagellum is a left-handed helix, the bacterium exhibits a clockwise circular motion. The radius of curvature, $r$, of this circular trajectory is given by
\begin{equation}
\begin{split}
&r=\frac{U_{\parallel}}{|W_{\perp}|}=\frac{\sqrt{U_{x}^{2}+U_{y}^{2}}}{|W_{z}-\left(W_{x}\cos\phi+W_{y}\sin\phi\right)\cot\Psi^{*}|},
\label{eq:refname19}
\end{split}
\end{equation}
where $U_{\parallel}$ is the translational velocity component parallel to the surface and $W_{\perp}$ is the rotational velocity component perpendicular to the surface (along the $z$-axis). A detailed expression for the radius of curvature of circular motion is provided in Appendix~\ref{appA}. The radius of curvature, $r$, which depends on the distance $l$ and the Debye length $\kappa^{-1}$, is shown in Fig.~\ref{fig:fig8}.

Fig.~\ref{fig:fig8}(a) shows that $r$ increases with increasing distance $l$. Similarly, Fig.~\ref{fig:fig8}(b) indicates that lower ionic strengths lead to a larger radius of curvature. The effect of motor rotation frequency on $r$ is negligible for the frequencies tested. In our simulations, the contour length of the flagellum is set to $\Lambda=7.5$~\si{\mu m}, consistent with the reported values of \emph{E. coli}~\cite{Darnton2007}. The radii of curvature of the circular motions fall within the range of several to tens of micrometers, similar to experimental values~\cite{Lauga2006,Dunstan2012}.

\section{Summary and Conclusions}
We simulate the dynamics of a bacterium swimming near a solid surface. The bacterium is simplified into a chiral two-body model, neglecting the hydrodynamic interactions between the flagellum and the solid surface. The entrapment process is characterized by height $h$ and inclination angle $\Psi$. This process is divided into three main stages: approach, reorientation, and surface swimming. Initially, the bacterium swims toward the surface at a velocity approximately proportional to the motor rotation frequency. During the reorientation stage, the cotangent of the inclination angle decays exponentially as a function of the product of the motor rotation frequency and time. The bacterium eventually reaches a stable fixed point $(h^{*},\Psi^{*})$ determined by the combined action of the near-field hydrodynamic interactions and the DLVO forces. Finally, the bacterium undergoes a stable circular motion on the surface. Bacteria with a left-handed flagellum exhibit a clockwise circular motion. The radii of curvature of these circular trajectories are closely related to the flagellar morphology and ionic strength of the electrolyte solution.

These results indicate that the stable height $h^{*}$ is insensitive to the contour length of the flagella and the motor rotation frequency. However, reducing the ionic strength increases the stable height, which can range from several nanometers up to over one hundred nanometers, broadly consistent with the experimental observations. This may explain the dispersion of stable heights observed experimentally for bacteria that swim near solid surfaces. Similarly, the stable inclination angle depends strongly on both the contour length and the ionic strength but is insensitive to the motor rotation frequency. Furthermore, an empirical power law approximately relates the stable inclination angle, $\Psi^{*}$, to the distance, $l$, between the centers of the cell body and the flagellum, given by $-\cos\Psi^{*}\propto l^{-2.5}$. The stable inclination angle decreases with increasing contour length or decreasing ionic strength. During stable circular motion, both longer flagellar contours and lower ionic strengths lead to larger radii of curvature $r$, typically ranging from several micrometers to tens of micrometers. Note that the zeta potential also affects the DLVO force. Therefore, the experimentally measured zeta potential values must be considered when validating the stable height, inclination angle, and radius of curvature of the circular motion.

\section*{Conflicts of interest}
There are no conflicts of interest to declare.


\begin{acknowledgments}
We acknowledge computational support from Beijing Computational Science Research Center and Ningxia University.
\end{acknowledgments}

\appendix
\section{Radius of Curvature of the Circular Trajectory}
\label{appA}
The Rodrigues' rotation matrix is given by the following formula~\cite{Dai2015,Murray2017,Liu2025B}
\begin{equation}
\begin{split}
\mathbf{R}(\hat{\mathbf{e}},\Delta\varphi)&=\cos\Delta\varphi\mathbb{I}+(1-\cos\Delta\varphi)\hat{\mathbf{e}}\otimes\hat{\mathbf{e}}+\sin\Delta\varphi\hat{\mathbf{e}}\times\\
&\approx\mathbb{I}+\Delta\varphi\hat{\mathbf{e}}\times,
\label{eq:refnameA01}
\end{split}
\end{equation}
where $\mathbb{I}$ is a $3\times3$ identity matrix, $\otimes$ is the Kronecker product, and $(\mathbf{e}\times)$ represents the $3\times3$ antisymmetric matrix such that $(\mathbf{e}\times)\mathbf{v}=\mathbf{e}\times\mathbf{v}$ for any vector $\mathbf{v}$. $\Delta\varphi=|\mathbf{W}_{b}|\Delta t$ is the rotation angle and $\mathbf{e}=\mathbf{W}_{b}/|\mathbf{W}_{b}|=(e_{x},e_{y},e_{z})$ is the unit vector along the axis of rotation. Once the bacterium reaches the stable fixed point $(h^{*},\Psi^{*})$, the initial direction of the flagellar axis is $\mathbf{e}_{1}=(0,\sin\Psi^{*},\cos\Psi^{*})$. After a time interval $\Delta t$, the direction of the flagellar axis becomes
\begin{equation}
\begin{split}
\mathbf{e}_{2}&=\mathbf{R}(\hat{\mathbf{e}},\Delta\varphi)\mathbf{e}_{1}=\left(e_{y}\Delta\varphi\cos\Psi^{*}-e_{z}\Delta\varphi\sin\Psi^{*},\right.\\ &\left.\sin\Psi^{*}-e_{x}\Delta\varphi\cos\Psi^{*},\cos\Psi^{*}+e_{x}\Delta\varphi\sin\Psi^{*}\right).
\label{eq:refnameA02}
\end{split}
\end{equation}
At the stable fixed point, the component of the rotational velocity of the cell body along the $x$-axis is zero, that is, $e_{x}=0$. The rotation angle $\Delta\phi$ of the bacterium around the $z$-axis is defined as the angle between the projections $\mathbf{e}_{1}$ and $\mathbf{e}_{2}$ onto the $xy$-plane, given by
\begin{equation}
\begin{split}
\cos\Delta\phi&=\frac{\sin\Psi^{*}}{\sqrt{(e_{y}\Delta\varphi\cos\Psi^{*}-e_{z}\Delta\varphi\sin\Psi^{*})^{2}+\sin^{2}\Psi^{*}}}\\
&=\frac{1}{\sqrt{1+(e_{y}\Delta\varphi\cot\Psi^{*}-e_{z}\Delta\varphi)^{2}}}.
\label{eq:refnameA03}
\end{split}
\end{equation}
Expanding both sides of the above equation using a Taylor series, retaining terms up to the second order:
\begin{equation}
\begin{split}
&1-\frac{\Delta\phi^{2}}{2}=1-\frac{(e_{y}\Delta\varphi\cot\Psi^{*}-e_{z}\Delta\varphi)^{2}}{2}.
\label{eq:refnameA04}
\end{split}
\end{equation}
The approximate value of the rotation angle $\Delta\phi$ can be expressed as follows:
\begin{equation}
\begin{split}
&|\Delta\phi|=|e_{y}\Delta\varphi\cot\Psi^{*}-e_{z}\Delta\varphi|.
\label{eq:refnameA05}
\end{split}
\end{equation}
Combining the following relationships:
\begin{equation}
\begin{split}
&W_{y}=e_{y}\Delta\varphi/\Delta t,\\
&W_{z}=e_{z}\Delta\varphi/\Delta t.
\label{eq:refnameA06}
\end{split}
\end{equation}
The radius of curvature, $r$, of the circular motion can be expressed as:
\begin{equation}
\begin{split}
&r=\frac{U_{\parallel}}{|W_{\perp}|}=\frac{|U_{y}|}{|W_{z}-W_{y}\cot\Psi^{*}|}.
\label{eq:refnameA07}
\end{split}
\end{equation}
Now consider the more general case in which the flagellar axis is $\mathbf{e}_{a}=(\sin\Psi^{*}\cos\phi,\sin\Psi^{*}\sin\phi,\cos\Psi^{*})$. The component of bacterial rotational velocity perpendicular to the solid surface, $W_{\perp}$, can be expressed as:
\begin{equation}
\begin{split}
&W_{\perp}=W_{z}-\left(W_{x}\cos\phi+W_{y}\sin\phi\right)\cot\Psi^{*}.
\label{eq:refnameA08}
\end{split}
\end{equation}
A more general formula for the radius of curvature is given by
\begin{equation}
\begin{split}
&r=\frac{U_{\parallel}}{|W_{\perp}|}=\frac{\sqrt{U_{x}^{2}+U_{y}^{2}}}{|W_{z}-\left(W_{x}\cos\phi+W_{y}\sin\phi\right)\cot\Psi^{*}|}.
\label{eq:refnameA09}
\end{split}
\end{equation}

\section{Resistance Matrix of a Sphere Near a Solid Surface}
\label{appB}
When a spherical cell body is close to a solid surface, the elements of its resistance matrix are~\cite{Dunstan2012}:
\begin{equation}
\begin{split}
\frac{Y_{\parallel}^{A}}{6\pi\mu R_{b}}=&\left(1.9963\xi-0.5332\right)\log\left(\frac{\xi}{\xi+1}\right)+2.9963\\
&-\frac{0.9689}{\xi+1}-\frac{0.5993}{(\xi+1)^{2}}-\frac{0.4691}{(\xi+1)^{3}},
\label{eq:refnameB01}
\end{split}
\end{equation}
\begin{equation}
\begin{split}
&\frac{Y_{\perp}^{A}}{6\pi\mu R_{b}}=\frac{2+9\xi+6\xi^{2}}{2\xi+6\xi^{2}},
\label{eq:refnameB02}
\end{split}
\end{equation}
\begin{equation}
\begin{split}
\frac{Y^{B}}{6\pi\mu R_{b}^{2}}=&\left(0.4991\xi+0.1334\right)\log\left(\frac{\xi}{\xi+1}\right)+0.4991\\
&-\frac{0.1162}{\xi+1}-\frac{0.0165}{(\xi+1)^{2}}+\frac{0.0028}{(\xi+1)^{3}},
\label{eq:refnameB03}
\end{split}
\end{equation}
\begin{equation}
\begin{split}
\frac{Y_{\parallel}^{C}}{8\pi\mu R_{b}^{3}}=&-\left(0.4+0.7898\xi\right)\log\left(\frac{\xi}{\xi+1}\right)+0.2101\\
&-\frac{0.0050}{\xi+1}-\frac{0.0683}{(\xi+1)^{2}}+\frac{0.2449}{(\xi+1)^{3}},
\label{eq:refnameB04}
\end{split}
\end{equation}
\begin{equation}
\begin{split}
\frac{Y_{\perp}^{C}}{8\pi\mu R_{b}^{3}}=&\left[0.414\xi+0.318\frac{\xi^{2}}{\xi+1}\right]\log\left(\frac{\xi}{\xi+1}\right)+1.732\\
&-\frac{0.684}{\xi+1}+\frac{0.037}{(\xi+1)^{2}}+\frac{0.117}{(\xi+1)^{3}},
\label{eq:refnameB05}
\end{split}
\end{equation}
where $\xi=h/R_{b}$, and $h$ is the minimum gap between the cell body and the solid surface.

\section{Resistance Matrix of the Flagellum}
\label{appC}
The bacterium is simplified into a chiral two-body model. The flagellum is represented by a single chiral body whose motion is described by a $6\times6$ resistance matrix~\cite{Di2011,Dvoriashyna2021,Liu2026A}. The detailed expressions for the components $X_{\parallel}^{A}$, $X_{\perp}^{A}$, $X_{\parallel}^{B}$, $X_{\perp}^{B}$, $X_{\parallel}^{C}$ and $X_{\perp}^{C}$ of this resistance matrix are given by:~\cite{Liu2026A}:
\begin{equation}
\begin{split}
X_{\parallel}^{A}=\Lambda\left[k_{\parallel}\cos^{2}\theta+k_{\perp}\sin^{2}\theta\right],
\label{eq:refnameC01}
\end{split}
\end{equation}
\begin{equation}
\begin{split}
X_{\perp}^{A}=\Lambda\left[k_{\parallel}\frac{\sin^{2}\theta}{2}+k_{\perp}\frac{1+\cos^{2}\theta}{2}\right],
\label{eq:refnameC02}
\end{split}
\end{equation}
\begin{equation}
\begin{split}
X_{\parallel}^{B}=RL\sin\theta\left(k_{\perp}-k_{\parallel}\right),
\label{eq:refnameC03}
\end{split}
\end{equation}
\begin{equation}
\begin{split}
X_{\perp}^{B}=-\frac{1}{2}RL\sin\theta\left(k_{\perp}-k_{\parallel}\right),
\label{eq:refnameC04}
\end{split}
\end{equation}
\begin{equation}
\begin{split}
X_{\parallel}^{C}=\Lambda R^{2}\left[k_{\parallel}\sin^{2}\theta+k_{\perp}\cos^{2}\theta\right],
\label{eq:refnameC05}
\end{split}
\end{equation}
\begin{equation}
\begin{split}
X_{\perp}^{C}=\Lambda\left[k_{\perp}(\frac{R^{2}}{2}+\frac{L^{2}}{12})+\left(k_{\parallel}-k_{\perp}\right)\sin^{2}\theta(\frac{R^{2}}{2\gamma^{2}}+\frac{L^{2}}{24})\right],
\label{eq:refnameC06}
\end{split}
\end{equation}
where $\Lambda$ is the contour length, $\theta$ is pitch angle, $R$ is helix radius, $\lambda$ is pitch, $L=\Lambda\cos\theta$ is axial length, and $\gamma=\tan\theta=2\pi R/\lambda$. The Gray and Hancock's drag coefficients are~\cite{Gray1955,Chwang1975}:
\begin{equation}
\begin{split}
&k_{\parallel}=\frac{2\pi\mu}{\ln(2\lambda/a)-1/2},\\
&k_{\perp}=\frac{4\pi\mu}{\ln(2\lambda/a)+1/2},
\label{eq:refnameC07}
\end{split}
\end{equation}
where $\mu$ is the dynamic viscosity of the fluid and $a$ is the filament radius of the flagellum.

\nocite{*}
\bibliography{aipsamp}

\end{document}